\documentclass[%
 reprint,
superscriptaddress,
nofootinbib,
 amsmath,amssymb,
pra,
longbibliography 
[]{revtex4-1}

\usepackage{amsmath,color}
\usepackage{graphicx}
\usepackage{dcolumn}
\usepackage{bm}
\usepackage{amsfonts}
\usepackage{amsthm}
\usepackage{comment}
\usepackage{ulem} 

\usepackage[usenames,dvipsnames]{xcolor}
\usepackage[colorlinks,citecolor=BlueViolet,linkcolor=RedOrange,urlcolor=BlueViolet]{hyperref}

\newcommand{\Emax}{E_{\textrm{max}}}
\newcommand{\Vmax}{V_{\textrm{max}}}
\newcommand{\sigmaexpsq}{\sigma_\textrm{exp}^2}
\newcommand{\gammaPhi}{\gamma_\Phi}
\newcommand{\DeltaTip}{\Delta_{\textrm{tip}}}
\newcommand{\DeltaSIC}{\Delta_{\textrm{SIC}}}
\newcommand{\GammaTip}{\Gamma_{\textrm{tip}}}
\newcommand{\GammaSIC}{\Gamma_{\textrm{SIC}}}

\newcommand{\COMPLETE}[1]{}

\newcommand{\CORR}[1]{{\color{blue} {#1}}}
\newcommand{\Igor}[1]{{\color{orange} {#1}}}
\newcommand{\pf}[1]{{\color{RedOrange}{#1}}}
\newcommand{\Mathieu}[1]{{\color{orange} {#1}}}
\newcommand{\referee}[1]{{\color{black} {#1}}}

\renewcommand{\sout}[1]{}
\renewcommand{\CORR}[1]{{{#1}}}
\renewcommand{\Igor}[1]{{ {#1}}}
\renewcommand{\Mathieu}[1]{{ {#1}}}
\renewcommand{\pf}[1]{{{#1}}}

\DeclareMathOperator{\re}{Re}

\begin{document}

\title{Local density of states fluctuations in 2D superconductor \\ as a probe of quantum diffusion}
\author{Mathieu Liz\'ee} 
\affiliation{Laboratoire de Physique de l'\'Ecole Normale Sup\'erieure, ENS, Universit\'e PSL, CNRS, 3 Sorbonne Universit\'e, Universit\'e Paris Cit\'e, 75005 Paris, France}
\affiliation{Sorbonne Universite, CNRS, Institut des Nanosciences de Paris, UMR7588, F-75252 Paris, France}

\author{Matthias Stosiek} 
\affiliation{Physics Division, Sophia University, Chiyoda-ku, Tokyo 102-8554, Japan}
\author{Christophe Brun}
\affiliation{Sorbonne Universite, CNRS, Institut des Nanosciences de Paris, UMR7588, F-75252 Paris, France}

\author{Igor Burmistrov}
\affiliation{\hbox{L.~D.~Landau Institute for Theoretical Physics, acad. Semenova av. 1-a, 142432 Chernogolovka, Russia}}
\affiliation{\hbox{Laboratory for Condensed Matter Physics, HSE University, 101000 Moscow, Russia}}

\author{Tristan Cren}
\affiliation{Sorbonne Universite, CNRS, Institut des Nanosciences de Paris, UMR7588, F-75252 Paris, France}

\begin{abstract}
The interplay of superconductivity and disorder generates a wealth of complex phenomena. In particular, the peculiar structure of diffusive electronic wavefunctions is predicted to increase the superconducting critical temperature in some range of disorder. 
In this work, we use an epitaxial monolayer of lead showing a simple band structure and homogenous structural disorder as a model system of a 
\Igor{2D} superconductor in the weak-antilocalization regime. \Mathieu{Then, we perform an extensive study of the emergent fluctuations of local density of states (LDOS) and spectral energy gap in this material and compare them with both analytical results and numerical solution of the attractive Hubbard model.} We show that mesoscopic LDOS fluctuations allow to probe locally both the elastic and inelastic scattering rates which are notoriously difficult to measure in transport measurements.
\end{abstract}

\maketitle

\section{Introduction}
Since the seminal paper by \Igor{P. W.} Anderson, the field of wave localization in disordered media has developed immensely. In the metallic regime, mesoscopic fluctuations of conductance stemming from the diffusion of electrons in a quenched disorder potential have been observed in a wealth of condensed matter systems and are commonly referred to as the `weak localization' signature \cite{Akkermans2007MesoscopicPO}. A similar signature of weak localization is predicted to emerge in maps of local density of states (LDOS) of 2D metallic systems \cite{wegner_inverse_1980,castellanits_multifractal_nodate,lerner_distribution_1988} and has already been observed for surface-plasmon modes \cite{carminati_electromagnetic_2015,krachmalnicoff_fluctuations_2010}.
For electronic modes however, despite several reports of electronic LDOS spatial fluctuations \cite{richardella_visualizing_2010, jack_visualizing_2021,zhao_disorder-induced_2019, morgenstern_real-space_2003}, theoretical predictions still lack a quantitative comparison with experiments.

The interplay of disorder and superconductivity has recently shown a renewed \Igor{experimental} (see Refs. \cite{gantmakher_superconductor-insulator_2010, sacepe_quantum_2020} for a review) 
\Igor{and theoretical} interest \cite{feigelman_fractal_2010,DellAnna,burmistrov_superconductor-insulator_2015,burmistrov_local_2016,Gastiasoro2018,Fan2020,stosiek_self-consistent-field_2020,Fan2020b,burmistrov_multifractally-enhanced_2021,stosiek_multifractal_2021,Burmistrov2022,Zhang2022}. In particular, the pairing of weakly localized `multifractal' electrons was surprisingly predicted to yield a disorder-enhanced $T_c$ compared to the clean metal case \CORR{in well chosen conditions} \cite{FeigelmanYuzbashyan2007, burmistrov_enhancement_2012}. \Igor{Recently,} a first \CORR{experimental demonstration of a possible multifractal enhancement of $T_c$} in NbSe$_2$ monolayers has been reported \cite{zhao_disorder-induced_2019}. Subsequently, the spatial distribution of the superconducting gap in this material has recently demonstrated multifractal statistics \cite{rubio-verdu_2020}. However, a clear picture of multifractal superconductivity is still lacking because the multifractal properties of the underlying eigenstates were not revealed in systematic LDOS measurements. In addition, a recent study suggests clearly that this experimental material presents inverse proximity effect from the graphene bilayer making it more an SN bilayer than a pure 2D single layer \cite{wander_these}. Thus, a deeper understanding of 2D diffusive superconductors in the multifractal regime is now required to strengthen this discovery and stimulate the engineering of multifractally enhanced superconductors.

In this study, we probe the mesoscopic fluctuations of LDOS in a purely two-dimensional weakly disordered superconductor with high resolution scanning tunneling spectroscopy (STS). In contrast to previous STS studies on thin films \cite{sacepe2008_disorder,Mondal2011_phase_fluct,Noat2013,Sherman201_effect_coulomb,sacepe_pseudogap_2010,sacepe_localization_2011} and NbSe2 monolayers \cite{zhao_disorder-induced_2019,rubio-verdu_2020}, we prove quantitatively that coherent electronic diffusion controls both the LDOS fluctuations close to the superconducting coherence \referee{peaks} and the spectral energy gap fluctuations. 
To generalize our interpretation, we compare our measurements with self-consistent solutions of the attractive Hubbard model on state of the art system size \cite{stosiek_self-consistent-field_2020,stosiek_multifractal_2021}. We demonstrate that the energy dependency of mesoscopic LDOS fluctuations allows to extract both the elastic and inelastic scattering rate of low energy single particle excitations and argue that LDOS spatial fluctuations constitute a valuable toolbox for the study of \Igor{2D} diffusive systems.

\begin{figure*} 
\centering
\includegraphics[width=18cm]{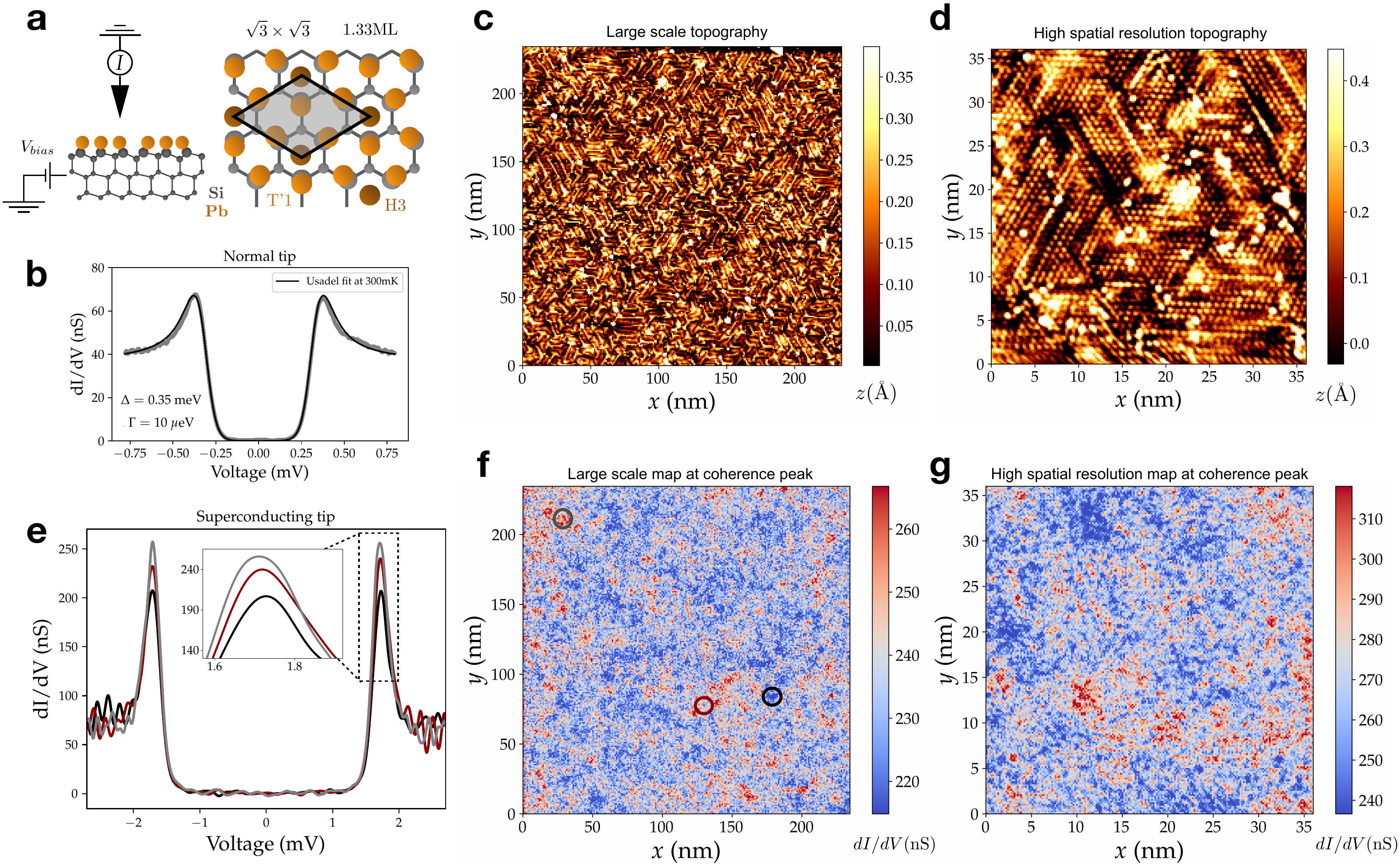} 
\caption{\textbf{Tunneling conductance fluctuations in the stripped incommensurate (SIC) phase of Pb/Si(111)} $\bold{a}$ Description of the scanning tunneling spectroscopy measurement and of the SIC phase (1.33 ML of Pb/Si with $\sqrt{3}\times\sqrt{3}$ symmetry). On \textbf{b}, we show the tunneling conductance of the SIC phase measured with a platinum tip along with the best fit with the Usadel formula (see Appendix \ref{Sec:AppA}). We find $\DeltaSIC=0.35$ meV and $\GammaSIC = 10$ $\mu$eV. \textbf{c} Topographic image of the SIC monolayer phase. One clearly sees the rotational nanodomains of the $\sqrt{3} \times \sqrt{3}$ phase separated by domain walls made of the $\sqrt{7} \times \sqrt{3}$ reconstruction. An atomically resolved image is shown in \textbf{d}. In $\bold{e}$,  several differential conductance spectra (dI/dV) measured at the positions shown on \textbf{f} are presented. On \textbf{f}, the isoenergy map corresponding to topography \textbf{c} at the coherence peak energy $\Vmax=1.7$meV is shown. Symmetrically, on \textbf{g}, we show the iso-energy dI/dV map corresponding to topography \textbf{d} at the same bias voltage.}
\label{Figure1}
\end{figure*}


\section{Experiments}
As model systems for the study of \Igor{\sout{two-dimensional} 2D} weakly disordered electronic systems, epitaxial monolayers of metals on semiconducting surfaces are exceptionally interesting. Firstly, their thickness of the order of the Fermi wavelength along with their good decoupling from the bulk makes them truly two-dimensional. Secondly, they show a wealth of phases with highly uniform and reproducible structural disorder without the need to evaporate chemical contaminants on the sample. Thus, these systems fully fabricated in ultra-high-vacuum are exceptionally clean and homogeneously disordered in contrast with the usually studied substitution alloys \cite{richardella_visualizing_2010,zhao_disorder-induced_2019,jack_visualizing_2021} where the disorder itself already has long range correlations as shown on topographic maps of references \cite{zhao_disorder-induced_2019,jack_visualizing_2021}.

In this work, we focus on the stripped incommensurate (SIC) phase of lead on silicon. The ideal SIC monolayer described in Figure~\ref{Figure1}\pf{\textbf{a}} is made of 1.33 monolayer of lead atoms on top of a Si(111) surface. The lead is evaporated on the 7$\times$7 reconstruction of silicon in a home made scanning tunneling microscope. The SIC phase is made of nanometric domains oriented preferentially along 3 directions as shown on large scale in \Igor{\sout{Figure 1\textbf{c,d}} Figures \ref{Figure1}\pf{\textbf{c}} and \ref{Figure1}\pf{\textbf{d}}}. The SIC phase has a nanometric mean free path much smaller than the superconducting coherence length $\xi \sim 50$ nm \cite{cherkez_proximity_2014} making it a prototype system to study \Igor{2D} diffusive superconductivity at weak disorder.

The sample is then cooled to 300 mK, well below its critical temperature of 1.8 K \cite{brun_review_2017}. \Igor{\sout{On panel \textbf{b}} In Figure \ref{Figure1}\pf{\textbf{b}}}, we show the tunneling spectroscopy at 300 mK measured with a platinum tip. The solid line is a solution of the Usadel equation for diffusive superconductors \cite{usadel_generalized_1970} (see Appendix \ref{Sec:AppA}). The spectrum of Figure~\ref{Figure1}\pf{\textbf{b}} was fitted using a gap of $\DeltaSIC = 0.35$ meV and a depairing energy of $\GammaSIC =$ 10 $\mu$eV.
In order to probe the mesoscopic fluctuations of this phase, we acquired several large scale (250 nm x 250 nm) spectroscopic maps with nanometric spatial resolution. A superconducting tip (bulk lead) is used to increase the energy resolution to $30$ $\mu$eV \cite{cherkez_proximity_2014,brun_remarkable_2014}. The $dI/dV$ spectrum (Figure \ref{Figure1}\pf{\textbf{e}}) shows sharp coherence peaks at $V_{\textrm{max}}\sim\Delta_{\textrm{tip}}+\DeltaSIC=1.7$ mV. Three individual spectra whose positions are shown in \Igor{Figure \ref{Figure1}\pf{\textbf{f}}} are displayed \Igor{in Figure \ref{Figure1}\pf{\textbf{e}}}.

Displaying the local differential conductance at a given bias voltage $V_\textrm{bias}$ yields iso-energy $dI/dV$ maps. On \Igor{\sout{Figure 1\textbf{f,g}} Figures \ref{Figure1}\pf{\textbf{f}} and \ref{Figure1}\pf{\textbf{g}}
}, we show the $dI/dV$ map measured at the energy of the coherence peak (1.7 mV) and evidence spatial fluctuations of the differential conductance spanning various scales below the superconducting coherence length $\xi \sim 50$ nm \cite{cherkez_proximity_2014,brun_remarkable_2014}. An interesting feature of this map is indeed that no characteristic scale can be easily identified. This \CORR{fractal-like behavior} is reminiscent of criticality close to the Anderson transition, \CORR{driven by the cooperation of disorder and electronic coherence.} 
The granular structure observed in topography \Igor{\sout{(panels \textbf{c,d})}(see Figures \ref{Figure1}\pf{\textbf{c}} and \ref{Figure1}\pf{\textbf{d}})} does not correlate at all with these `emergent' fluctuations that we consequently attribute to coherent diffusion.

\section{Analysis of fluctuations}

\begin{figure*}
\centering
\includegraphics[width=18cm]{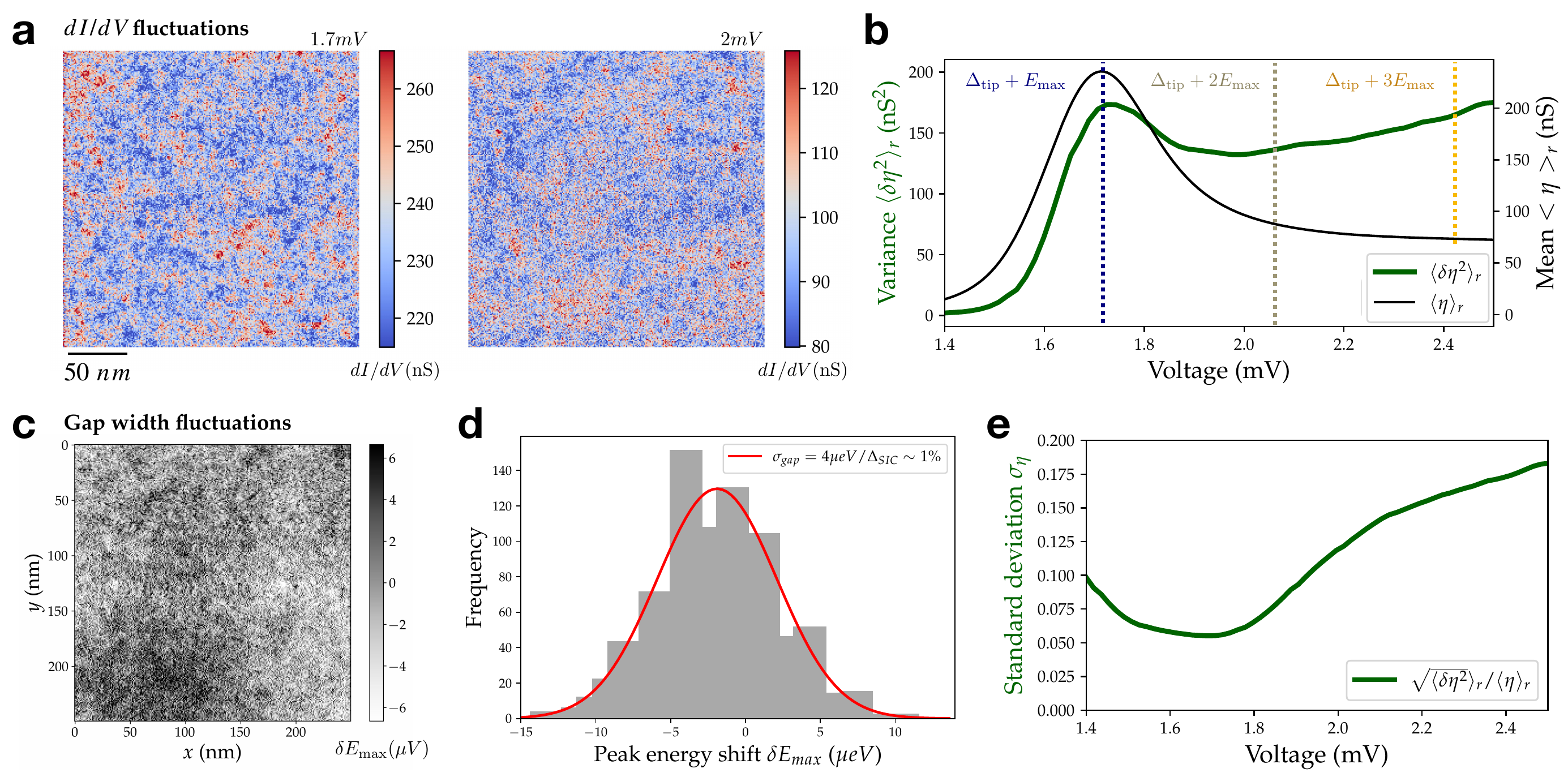} 
\caption{\textbf{Analysis of tunneling conductance fluctuations} \textbf{a} Iso-energy maps of the differential conductance $\eta = dI/dV$. \textbf{b} Variance (green) and mean value (black) of the tunneling conductance computed on iso-energy maps as a function of the bias voltage. \Mathieu{On \textbf{c}, show the map of the coherence peak energy fluctuations, corresponding to gap width fluctuations. On panel \textbf{d}, the distribution of this map is given along with a gaussian distribution of standard deviation $\Sigma_{gap} = 4\mu eV$. On \textbf{e}, we show the bias voltage dependency of the normalized standard deviation of dI/dV : $\sigma_\textrm{exp}=\sqrt{\langle \delta \eta(V)^2\rangle}/\langle \eta(V) \rangle $.}}
\label{Figure2}
\end{figure*}

In order to reveal superconducting properties in more details, we focus on LDOS spatial variance close to the superconducting coherence \referee{peaks}. At voltage $V$, the variance of iso-energy tunneling conductance maps (shown on \Igor{\sout{Figure 2\textbf{a}} Figure~\ref{Figure2}\pf{\textbf{a}}} for $V\in\{1.7,2\} $ mV) writes $\sigmaexpsq(V) = \langle\delta \left(\frac{dI}{dV}\right)^2\rangle_r$ where to keep notations light, \CORR{$\eta=dI/dV$ is the experimentally measured differential conductance} \Igor{(see Sec. \ref{Sec:Comp:Exp} for more accurate definition)} and the brackets denote spatial averaging.
In Figure \ref{Figure2}\pf{\textbf{b}}, we plot (in green) the variance $\langle \delta \eta^2 \rangle$ as a function of bias voltage, showing a maximum close to the coherence peak energy $\Vmax=\Emax+\DeltaTip = $1.7 mV. The normalized standard deviation is plotted in panel \textbf{e} and shows a characteristic minimum close to $\Emax$ followed by a convex increase at higher energy. \referee{As shown in Appendix \ref{Sec:AppE}, the normalized standard deviation of LDOS is symmetric with respect to the Fermi level (\textit{ie} : negative and positive bias voltage).} As gap width granularity is a standard feature of 2D superconductors \cite{sacepe_quantum_2020,rubio-verdu_2020,ghosal_inhomogeneous_2001,zhao_disorder-induced_2019}, we plot the fluctuations of the peak energy $\Emax$ on Figure~\ref{Figure2}\pf{\textbf{c}}. The distribution (\textit{cf:} panel \textbf{d}) shows a relative standard deviation $\sigma_{\Emax}$ of about 1$\%$, much smaller than the relative fluctuations of LDOS shown on panel \textbf{e} ranging from 6 to 20$\%$. 
\\

\subsection{Semi-analytical theoretical predictions \label{Sec:Sub:Theory}}

\begin{figure*}
\centering
\includegraphics[width=18cm]{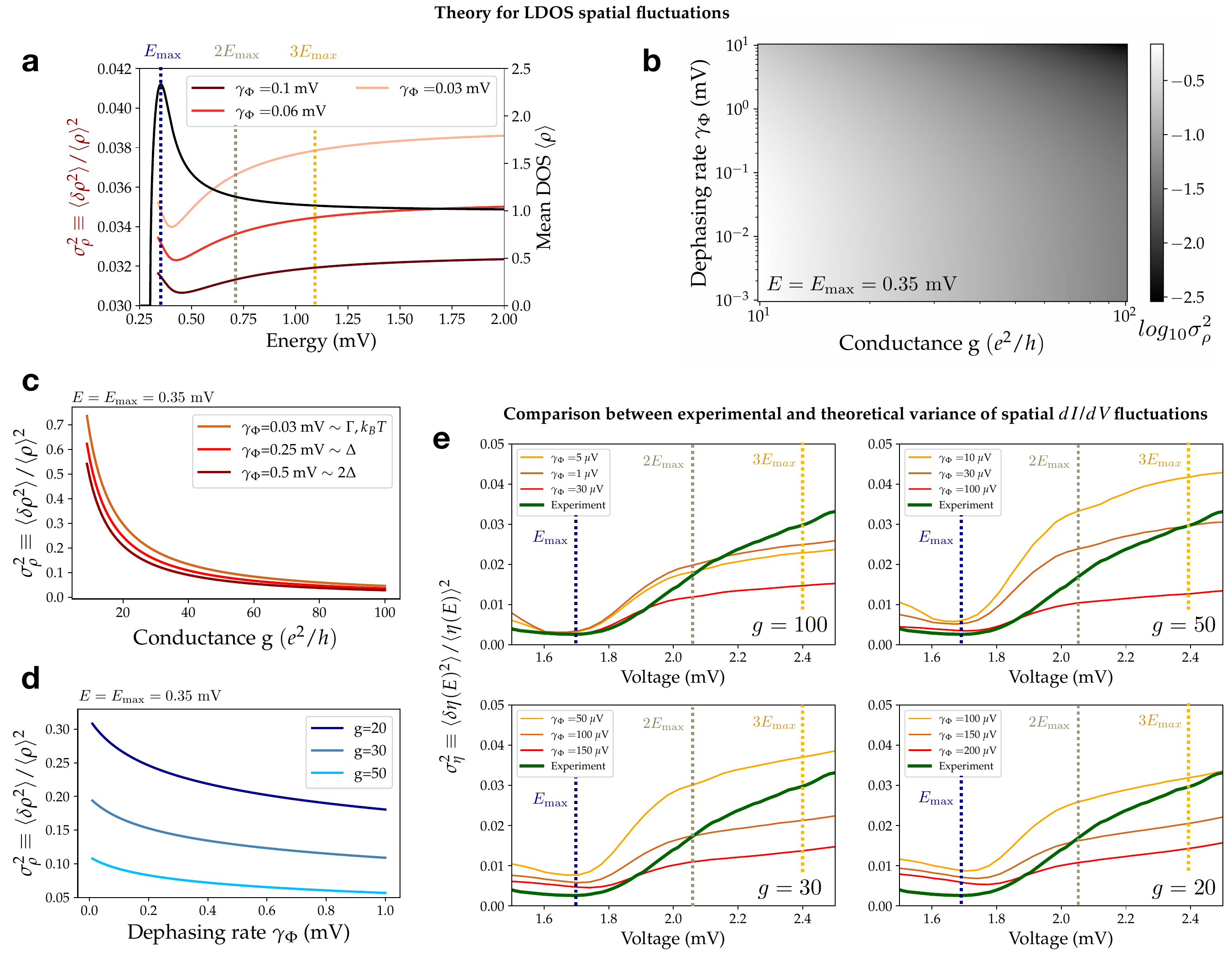} 
\caption{\textbf{Analysis of tunneling conductance fluctuations} On panels \textbf{a-d}, we describe the fluctuations of density of states computed semi-analytically as a function of energy E, conductance g and effective dephasing rate $\gamma_\Phi$. On \textbf{a}, we show the energy dependency of the normalized variance $\sigma_\rho^2$ (red) at $g=30$ and several values of $\gamma_\Phi \in \left\{0.03,0.06,0.1\right\} mV$. In black, we show the mean density of states used for the calculation (solution of Usadel equation, \textit{cf.} Appendix \ref{Sec:AppA}). On \textbf{b-d}, we now fix the energy at $\Emax$=0.35 mV and plot on \textbf{b} a color-map oftay the log of $\sigma_\rho^2$ as a function of both conductance and dephasing rate. On \textbf{c}, we show the dependency of $\sigma_\rho^2$ on conductance for $\gamma_\Phi \in \left\{0.03,0.25,0.5 \right\}$ mV. On \textbf{d}, we show the dependency on dephasing rate for g$\in \left\{20,30,50\right\}$. On \textbf{e}, we compare the experimentally measured normalized variance of tunneling conductance $\sigma_\eta^2$ with the one computed semi-analytically from density of states correlations (Eq.\eqref{appendixB_dos} derived in Appendix \ref{Sec:AppB} from Eq.\eqref{LDOS_corr_main_text}). We plot the theoretical prediction for $g\in\left\{20,30,50,100\right\} $ and the levels of dephasing which best reproduce the experimental data in each case. We stress that no free parameter is used here.}
\label{Figure3}
\end{figure*}

To rationalize the energy dependency of LDOS spatial variance, we now compute the fluctuations of density of states\Igor{, $\rho(E,\bm{r})$,} in a 
\Igor{2D} diffusive superconductor. In the following, we sketch a simplified derivation\Igor{\sout{of the density of states correlator in diffusive superconductors}. We have $\rho(E,r)=-\frac{1}{\pi}$Im $G^R (E,r,r)$ with $G^R (E,r,r)=\langle r|\frac{1}{E-H+i\eta}|r \rangle$.} Like in the mean-field solution to the BCS Hamiltonian, we introduce the Bogoliubov operators and coherence factors \Igor{
$u^2_\alpha/v^2_\alpha=(1\pm\epsilon_\alpha/\sqrt{\epsilon_a^2+\Delta^2})/2$
}
associated to the single particle eigenstates $\phi_\alpha$ for the eigenvalue $\epsilon_\alpha$, solutions of the single particle Schrödinger equation including the disordered potential. The LDOS can be written conveniently as:
\begin{equation}
\rho(E,\bm{r})=\sum_{\alpha,s=\pm}\phi_\alpha^2(\bm{r})(1+\epsilon_\alpha/E)\delta(E-s\sqrt{\epsilon_\alpha^2+\Delta^2}) .
\end{equation}
Here $s=\pm$ denotes states respectively above of below the Fermi level.
The density of states correlations are then computed from this expression along with the dynamical structure factor which is a spatially averaged product of wavefunctions measured at position $\bm{r}$ and $\bm{r^\prime}$ and at well determined energies. If one neglects the dependency of single particle density of states on energy \CORR{in the normal state}, this structure factor can be computed from the polarization operator $\Pi^R(\omega,\bm{r},\bm{r^\prime})$ \Igor{(see, e.g. \cite{lee_disordered_1985})}
\begin{equation}
    F(\epsilon,\omega,\bm{r},\bm{r^\prime})\simeq\frac{1}{\pi\omega}{\rm Im}\, \Pi^{\rm R}(\omega,\bm{r},\bm{r^\prime}) .
\end{equation}
In the \textit{diffusive regime at weak disorder}, one gets for the Fourier transform of the polarization operator
\begin{equation}
    \Pi^R(\omega,\bm{q})=\rho_0\frac{Dq^2}{Dq^2-i\omega} ,
\end{equation}
where $D$ denotes the diffusion coefficient and $\rho_0$ the density of states \CORR{of the noninteracting problem at the Fermi level}. The full derivation then yields the pair LDOS correlation function at different \Igor{spatial} points ($\bm{r_1}$,$\bm{r_2}$) and different energies ($E_1$,$E_2$) \CORR{\cite{burmistrov_multifractally-enhanced_2021,stosiek_multifractal_2021,Burmistrov2022}}:
\begin{widetext}
\begin{eqnarray}
\langle\delta\rho(E_1,\bm{r}_1)\delta\rho(E_2,\bm{r}_2)\rangle = \frac{\rho_0^2}{2\pi g}
\re \Biggl\{\Bigl[1+X_{E_1}X_{E_2}^*\Bigr] K_0\left (R\sqrt{\frac{2\gamma_{\Phi}-iE_1/X_{E_1}+iE_2/X_{E_2}^*}{\hbar D}}\right) \notag \\
-  \Bigl[1-X_{E_1}X_{E_2}\Bigr] K_0\left (R\sqrt{\frac{2\gamma_{\Phi}-iE_1/X_{E_1}-iE_2/X_{E_2}}{\hbar D}} \right)\Biggr\} ,
\label{LDOS_corr_main_text}
\end{eqnarray}
\end{widetext}
where $R=|\bm{r}_1-\bm{r}_2|$, $K_0(z)$ stands for the modified Bessel function. Also, we introduce $X_E = E/\sqrt{E^2-\Delta_E^2}$ where we phenomenologically substitute $\Delta$ by the complex energy dependent gap-function $\Delta_E$ which we estimate using the Usadel model for diffusive superconductors (see Appendix \ref{Sec:AppA}). \referee{We check that, as observed experimentally, Eq.\ref{LDOS_corr_main_text} is symmetric with respect to the Fermi level : $E_1 = E_2 =V$ and  $E_1 = E_2 =-V$ yield the same correlations.} 
Finally, two parameters control the strength of LDOS fluctuations: the dimensionless conductivity $g=h D \rho_0$/2 and the effective dephasing rate $\gammaPhi = \hbar D/L_\Phi^2$ which we assume to be energy independent. \referee{We note that strong spin-orbit coupling results in a factor $1/4$ (in comparison with the case when spin-orbit coupling is absent) due to suppression of triplet diffusons. This well-known fact, see e.g. Ref. \cite{altshuler1985electron}, was taken explicitly into account in Eq. \ref{LDOS_corr_main_text}, see Ref. \cite{Burmistrov2022}.} In Figure~\ref{Figure3}\pf{\textbf{a}}, we show the energy dependency of the density of states normalized variance $\sigma_\rho^2(E) = \langle \delta \rho(E)^2\rangle_r/\langle\rho(E)\rangle_r^2$. Here, the conductance is fixed at $g=30$ and we show the plots for several values of $\gamma_\Phi$ ranging typically between $\Gamma \sim k_B T \sim 30 \mu\textrm{V}$  and $\Delta \sim 350 \mu\textrm{V}$. In black, we show the mean density of states spectrum where $\re(X_E)$ used in this computation corresponds to the best Usadel fit for the SIC phase (see Appendix \ref{Sec:AppA}). We show that $\sigma_\rho ^2$ has a local minimum close to the coherence peak energy $\Emax$, in good agreement with the minimum of normalized variance $\sigmaexpsq$ at the superconducting coherence peak (Figure~\ref{Figure2}\pf{\textbf{c}}). \Igor{\sout{On panels \textbf{b,c}} In Figures~\ref{Figure3}\pf{\textbf{b}} and \ref{Figure3}\pf{\textbf{c}}}, we show that at fixed energy, $\sigma_\rho^2$ decreases with increasing conductivity (as expected from the $1/g$ dependency in equation \eqref{LDOS_corr_main_text}). \Igor{\sout{On panels \textbf{b,d}} In Figures~\ref{Figure3}\pf{\textbf{b}} and \ref{Figure3}\pf{\textbf{d}}}, we \Igor{\sout{show} demonstrate} that increasing dephasing rate $\gamma_\Phi$ -- corresponding to smaller system size in a transport experiment (Thouless energy) -- \Igor{\sout{decreases} reduces} the variance $\sigma_\rho^2$. We stress that these dependencies are very natural in the context of mesoscopic fluctuations. Larger electronic phase coherence length and stronger disorder lead to enhanced fluctuations, whether one measures conductance in transport experiments or LDOS with a STM. 

\begin{figure*} 
\centering
\includegraphics[width=18cm]{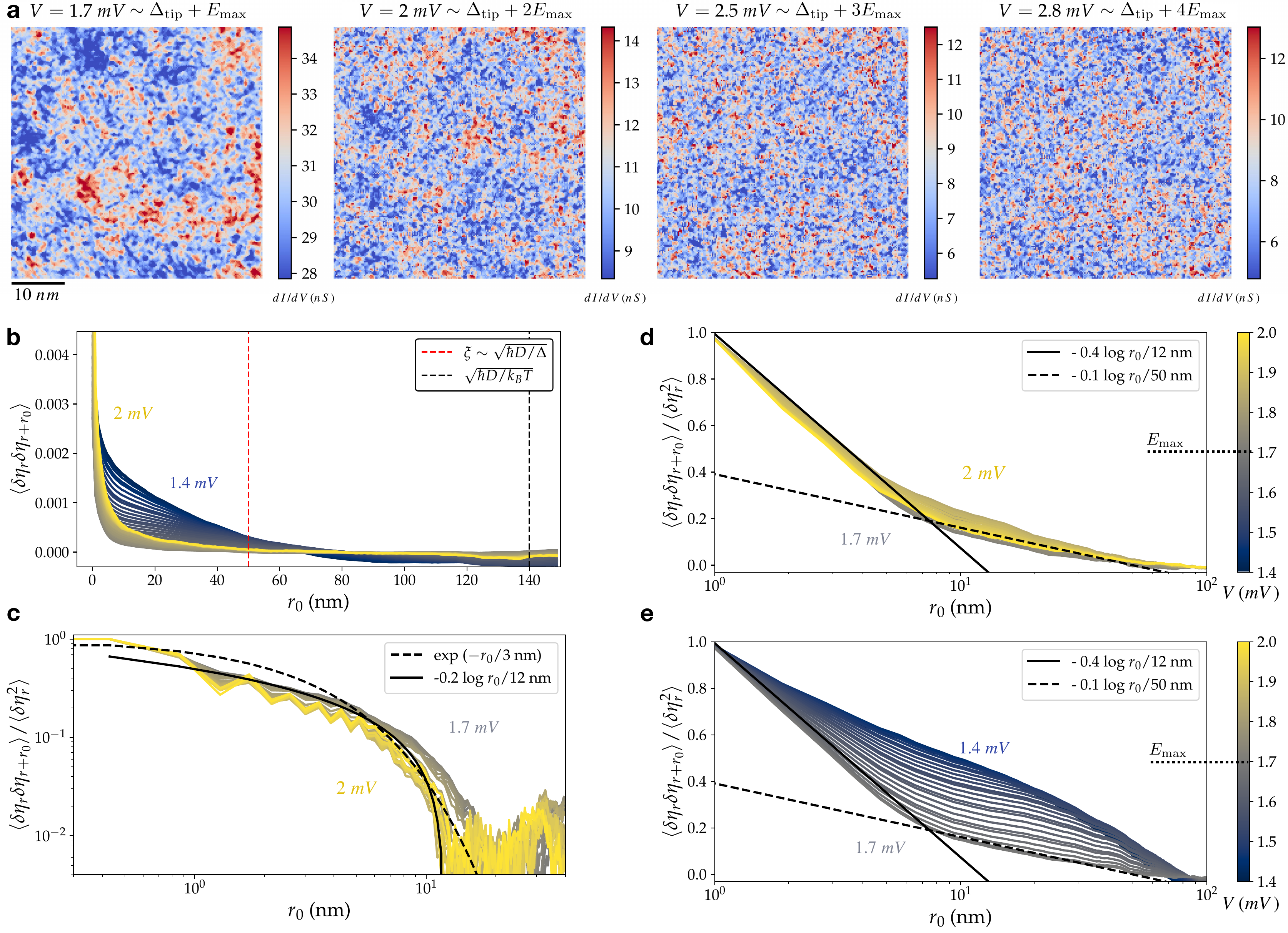} 
\caption{\textbf{dI/dV spatial correlations} On \textbf{a}, we show several iso-energy high resolution differential conductance maps for $E = V-\DeltaTip \in \left\{1,2,3,4\right\} \Emax$. On panel \textbf{b}, we show the angle averaged 2-point correlations on large scale maps as a function of distance and bias voltage (color scale for panels \textbf{b-e} given on panels \textbf{d,e}). On this plot, we also show the position of superconducting coherence length $\xi$ (red) and the diffusion length corresponding to energy scale  $k_BT \sim 30$ $\mu$eV (black). On \textbf{c}, we plot the angle averaged auto-correlation computed on small scale maps. On \textbf{d,e}, we use lin-log scales to evidence the peculiar behavior of the auto-correlation (normalized by its zero distance value). On \textbf{d}, we focus on energies below $\Emax$ and on \textbf{e}, on energies above $\Emax$.}
\label{Figure4}
\end{figure*}

 \subsection{Electronic diffusion in the SIC phase and quantitative extraction of transport parameters}
 
 \CORR{In order to compare the amplitude of LDOS fluctuations with theoretical predictions, we evaluate the diffusion coefficient in the SIC phase by independent means.} We refer to previous work by some of the authors where the proximity effect between the SIC phase and small bulky lead islands allowed to estimate the diffusion coefficient of the monolayer \cite{cherkez_proximity_2014}. These results are supported by another measurement where the spatial profile of a vortex core allowed us to extract the effective coherence length in the diffusive limit and thus the electronic diffusion coefficient \cite{brun_remarkable_2014}. Both these measurements yield a dirty coherence length $\xi \in$ [45,50] nm. Writing $\xi = \sqrt{{\hbar D}/{\Delta}\Igor{_{\textrm{SIC}}}}$ with $\Delta_{\textrm{SIC}}$=0.35 meV gives a diffusion coefficient $D \in$ [10,15] cm$^2$/s. We now consider the Einstein relation for the conductivity (per spin orientation) of the monolayer $g=h D \rho_0/2$ and using the 2D electron gas model \CORR{- shown to be appropriate by ARPES measurements \cite{choi2007}} - 
 we write for the density of states $\rho_0 = 2\frac{k_F}{h v_F}=\frac{2m}{2 \pi \hbar^2}$ (factor of 2 for the two spin orientations). This leaves us with $g=\frac{m D}{\hbar}$. Using the known effective mass $m=1.27 m_e$ in the SIC phase \cite{brun_review_2017}, we can estimate $g\sim$ 20. \CORR{We insist that this value may be under-estimated as in the former experiments used to evaluate $\xi$, the disorder was higher than in the present experiment due to scattering nano-islands not appearing here. Thus we infer that $g=20$ is in fact a lower bound of the actual conductance and we roughly estimate $g\in[20-100]$.
 Like we stated earlier, the lead monolayer is weakly disordered and lies deep in the diffusive regime as the mean free path $\ell \sim 1-5$ nm is much smaller than $\xi$.}
 
\Mathieu{\subsection{Gap width fluctuations \label{Subsec:gap_fluct_theory}}}
\Mathieu{For a diffusive 2D superconductor in the weak disorder regime, fluctuations of $\Emax$ can be estimated theoretically (\textit{cf}: Appendix \ref{Sec:AppD}). We find 
$\sigma_{\Emax} {\simeq} 
\sqrt{c/g} (\Gamma_{SIC}/\Delta_{SIC})^{2/3}$  where $c\sim 0.3$. 
Here again, these mesoscopic fluctuations scale like $1/\sqrt{g}$ and are extremely small compared to the peak width at weak disorder. In the relevant conductance range for our system $g\in[20-100]$, we obtain $\sigma_{\Emax} \in [0.5, 1.1]\%$, in excellent agreement with the 1\% obtained experimentally (\textit{cf} : Figure \ref{Figure2}\textbf{d}). We compare these results for a weakly disordered phase with what was obtained in much more disordered niobium nitride (NbN) thin films ($g\sim 2$) where the relative gap fluctuations are of the order of $6\%$ \cite{carbillet_spectroscopic_2020} in good agreement with the theory ($7\%$) and much higher than what we measure here.}


\subsection{LDOS variance : Comparison with experiments\label{Sec:Comp:Exp}}

We now attempt a quantitative comparison of our theoretical analysis \Mathieu{for the energy-dependent LDOS fluctuations} with our experiments on the SIC phase of lead on silicon. Using equation \eqref{LDOS_corr_main_text}, we compute the normalized variance of the tunneling conductance: $\sigma_\eta^2 = \langle \delta \eta(V)^2\rangle_r/\langle\eta(V)\rangle_r^2$ at a bias voltage V. Here, we take into account the tip density of states and variations of the tip height above the sample during the measurement. All details are given in Appendix \ref{Sec:AppB}. The formula we obtain for $\sigma_\eta^2$ (Eq. \eqref{appendixB_dos}) is a \Igor{\sout{simple} straightforward} energy integration of the density of states correlator given in equation \eqref{LDOS_corr_main_text}.
 
 
 The tunneling conductance variance can then be compared to the experimentally measured $\sigmaexpsq$ (in green) on Figure \ref{Figure3}\pf{\textbf{e}}. The thin lines are semi-analytical calculations of $\sigma_\eta$ for $g\in \left\{20,30,50,100\right\}$ from Eq. \eqref{appendixB_dos}. For each value of $g$, we plot the theoretical curve for a few dephasing rates $\gamma_\Phi$ which best reproduce the experimental variance. We stress that no free parameter is used here as $g$ takes very reasonable values for the SIC phase \cite{cherkez_proximity_2014,brun_remarkable_2014} while the dephasing rate $\gamma_\Phi$ is constrained \CORR{in a rather narrow window of physically relevant energies $\gammaPhi \in [1, 150]$ $\mu$eV $\sim  [k_B T/30, \DeltaSIC/2]$ . We obtain an excellent quantitative agreement between experiment and theoretical predictions for a range of parameters corresponding to $g\in [50, 100]$ where the energy dependency of the normalized variance close to the coherence peak is very nicely reproduced.}

Thus, we claim that emergent $dI/dV$ spatial fluctuations in the \Igor{2D} superconductor are a direct probe of coherent diffusion, well reproduced by a simple analytical model (section \ref{Sec:Sub:Theory}). 
Although a quantitative extraction of $\gammaPhi$ is difficult in our case, our results show that scanning tunneling spectroscopy intrinsically allows for a local measurement of both dimensionless conductance and electronic coherence length. This non invasive local probe explores structurally optimized regions far from step edges or contacts. This direct probe of electronic diffusion, should be considered as complementary to transport measurements. Moreover, it shows a huge - yet almost unexplored - potential for improvement through the detailed analysis of spatial correlations and in particular of the spatial structure of wavefunctions which is expected to be the very cause of multifractal $T_c$ enhancement. In this regard, performing constant-height spectroscopic maps would allow a more direct comparison to theoretical predictions and is therefore a very exciting perspective.

\section{Spatial correlations}

After considering the variance of the iso-energy maps and thus disregarding the spatial structure of $dI/dV$ fluctuations, we briefly focus on the analysis of spatial correlations of the local conductance: $\langle\delta \eta_r\delta \eta_{r+r_0}\rangle_{r}$. On Figure~\ref{Figure4}\pf{\textbf{a}}, we show several high resolution conductance maps at various energies above the coherence peak $E = V-\DeltaTip$ $\in \left\{0.35, 0.7, 1.2, 1.5\right\}$ $\textrm{mV}$ $\sim \left\{1,2,3,4\right\}\Emax$. We observe very clearly that large scale spatial structures at $\Emax$ tend to disappear with increasing energy.

To explore the iso-energy spatial correlations of tunneling conductance, we plot on Figure \ref{Figure4}\pf{\textbf{b}} the angle-averaged 2-point correlation function as a function of distance $r_0$. The color of the curve represents the energy at which it is measured. On Figure \ref{Figure4}\pf{\textbf{c}}, we use a log-log representation of the auto-correlation function on high resolution maps (normalized by its value at 0.1 nm). To better show the angle averaged radial decay at various energies on large scale 250 nm maps, we normalize the auto-correlation by its value at $r_0=1$ nm and plot them for energies below $\Emax$ on Figure \ref{Figure4}\pf{\textbf{d}} and above $\Emax$ on Figure \ref{Figure4}\pf{\textbf{e}} where the curve's color is coding its energy below 1.4 and 2 mV. It is apparent here that above $\Emax$, these auto-correlation profiles depend only very weakly on energy. On Figure \ref{Figure4}\pf{\textbf{d}}, we show that the spatial auto-correlation \Igor{function} has a short range regime with steep decay up to approximately 10 nm followed by a long range regime with a slower decay. As made visible with the black lines, two apparently log-decay regimes are identified with characteristic distance slopes of 12 and 50 nm for the short and long range regimes respectively. Although they are not true correlation lengths, these two characteristic lengths in the auto-correlation function seem very natural as 10 nm correspond to the typical nanocrystal size of the SIC monolayer and 50 nm to its superconducting coherence length.

To deepen our understanding of these LDOS fluctuations, we complement our experimental work with a numerical study of superconducting electrons on a 2D disordered lattice in the weak disorder regime. The deep similarity between experimental and \CORR{our fully self-consistent tight-binding models} then allow us to make very precise comparison between the numerical, experimental and analytical studies.

\section{Numerical study of LDOS fluctuations with the attractive Hubbard model}

\CORR{We write and solve a tight-binding model tailored to match the experimental system: a weakly disordered \Igor{2D} diffusive superconductor of comparable} conductivity and dephasing rates. We consider the attractive$-U$ Hubbard model on the square lattice in two-dimensions with double-periodic boundary conditions. Within the mean-field approximation the Hamiltonian reads ($U>0$)
\begin{gather}
\label{e6}
\hat H = 
-t\sum_{\langle i,j\rangle, \sigma} \hat{c}_{i,\sigma}^\dagger \hat{c}_{j,\sigma} + 
\sum_{i, \sigma} \bigl ( V_i-\mu-U n(\bm{r}_i)/2  \bigr) \hat{n}_{i,\sigma} 
\notag
\\ 
+ \sum_{i} \Delta(\bm{r}_i) \hat{c}_{i,\uparrow} \hat{c}_{i,\downarrow} + \text{h.c.}  
\label{eq:Ham:def}
\end{gather}
where $\hat{c}_{j,\sigma}^\dagger$ and  $\hat{c}_{j,\sigma}$ denote the creation and annihilation operators of an electron with spin $\sigma=\pm 1/2$ on site $j$. The on-site disorder potential is drawn from a box distribution, $V_i \in \left[-W,W \right]$ with a disorder strength fixed at W=0.5 in an attempt to match the experimental disorder strength. The chemical potential $\mu$  fixes the filling factor to $0.3$. Throughout this work the interaction is taken as $U=2.2 t$ and the system size as $L=192$, the local occupation number $n(\mathbf{r}_i)$ and the pairing amplitude $\Delta(\mathbf{r}_i)$ are determined self-consistently,
\begin{equation}
n(\textbf{r}_i)= \sum_\sigma \langle \hat{n}_{i,\sigma} \rangle, \qquad  \Delta(\textbf{r}_i)=U \langle \hat{c}^\dagger_{i,\downarrow} \hat{c}^\dagger_{i,\uparrow} \rangle , 
\label{eq:self-consistent}
\end{equation}
where $\hat{n}_{i,\sigma}=\hat{c}_{i,\sigma}^\dagger \hat{c}_{i,\sigma}$ \CORR{\sout{and the average is taken with respect to the equilibrium density matrix corresponding to the Hamiltonian \eqref{eq:Ham:def} at zero temperature}}. We solve Eqs.
\eqref{eq:Ham:def}--\eqref{eq:self-consistent} iteratively until a self-consistent solution is obtained (see Ref. \cite{stosiek_self-consistent-field_2020} for further computational details). The ensemble averaging involves typically more than a hundred samples and the density of states is computed by averaging on an energy scale $\langle\Delta\rangle/10$ in good matching with the experimental situation ($\GammaTip = 20$ $\mu$eV $\sim \DeltaSIC/10$). \referee{We note that the numerical model \label{eq:Ham:def} does not reproduce the strong spin-orbit coupling of the SIC phase. Nevertheless, as explained in section \ref{Sec:Sub:Theory}, the theoretical analysis predicts that strong S-O coupling reduces the normalized standard deviation $\sigma$ by a factor of 2 \cite{Burmistrov2022}. Keeping this two-fold reduction in mind allows one to quantitatively compare experiments, analytical predictions and numerical calculations.}

On Figure \ref{Figure5}, we show the results of the numerical investigation and compare them with both the analytical theory derived earlier and the experimental results. We check that, as expected at weak disorder \cite{Burmistrov2022} and in quantitative agreement with the experiment, the spectral gap shows very small fluctuations of about 2$\%$. On panels \ref{Figure5}\pf{\textbf{a}} and \ref{Figure5}\pf{\textbf{b}}, we show local density of state maps taken at $E = \Emax$ and $E=2 \Emax$. In excellent agreement with the experimental results (Figure \ref{Figure4}), we observe that LDOS fluctuations exhibit much longer range correlations close to $\Emax$ than at higher energy (\textit{cf:} size of structures in Figure \ref{Figure5}\textbf{a-b}).

\Mathieu{\subsection{Gap width fluctuations \label{subsect:gap_width_numerics}}}
\Mathieu{Following the analysis of \ref{Subsec:gap_fluct_theory} for the analytical theory, we start our systematic comparison between the SIC phase and the attractive Hubbard model by gap fluctuations. In the self-consistent numerical model, we find relative fluctuations $\sigma_{\Emax} \sim 2\%$ slightly above our experimental result of 1\% (\textit{cf:} Figure \ref{Figure2}\textbf{c}). A detailed study of the spectral gap and order parameter statistics in the disordered attractive Hubbard model is in preparation \cite{stosiek_gap}.}
\\
\subsection{Normalized variance}

Like in the experimental section above, we now proceed to study the normalized variance of the LDOS spatial fluctuations. On \Igor{\sout{panel \textrm{c}} Figure \ref{Figure5}\pf{\textbf{c}}}, we show the mean density of states of the numerical model (black line) along with the normalized variance $\sigma^2$ (red). We compare the numerically obtained LDOS variance with the analytical prediction derived from Eq. \eqref{LDOS_corr_main_text} depending on dimensionless conductance $g$ and dephasing rate $\gammaPhi$.  As our tight binding approach considers solutions of the stationary Schrödinger equation, it does not include dephasing. $\gammaPhi$ is thus \Igor{substituted by} a Thouless energy corresponding to diffusive motion at the scale of system size: $\gamma_\Phi =\frac{\hbar D}{L^2}$. Knowing the density of states and Fermi velocity of the \Igor{2D} electron gas, we manage to compute Eq.\eqref{LDOS_corr_main_text} as a function of a single parameter, the dimensionless conductance $g$ (with $\gammaPhi = gta^2/L^2$). \Mathieu{We find that $g\sim10$ reproduce almost perfectly the numerical results (red) on all the energy range we considered. This lower conductance compared to the SIC phase is fully consistent with the gap width fluctuations - twice higher for the model than for experiments. We conclude, based on enhanced fluctuations of both the LDOS and the gap width, along with a similar energy dependency that the tight binding system is slightly more disordered than its experimental counterpart.}

\subsection{LDOS distributions}

We already know from several experiments that disordered systems tend to yield log-normal LDOS distributions \cite{richardella_visualizing_2010, jack_visualizing_2021}. More precisely, close to Anderson transition, a log-normal distribution of the LDOS is expected \CORR{\cite{burmistrov_multifractally-enhanced_2021, lerner_distribution_1988, burmistrov_local_2016}}. In the low disorder regime relevant here, log normal distributions for the LDOS have also been predicted analytically \cite{falko_1995} and observed in numerical models \cite{stosiek_multifractal_2021}.

\begin{figure*} 
\centering
\includegraphics[width=18cm]{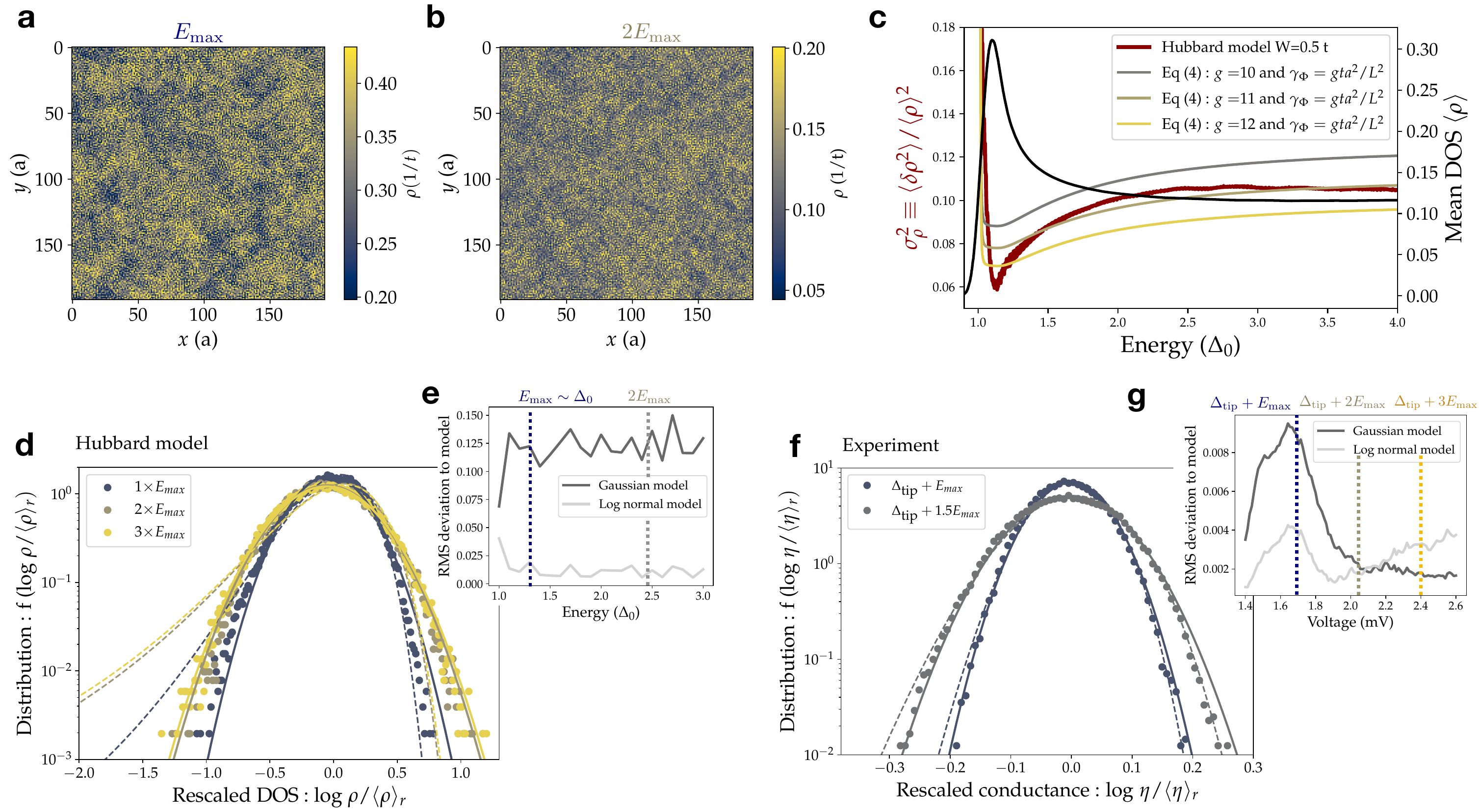} 
\caption{\textbf{Disordered attractive Hubbard model and LDOS distributions} On panel \textbf{a,b}, we show local density of state maps taken at E =$\Emax$ and 2$\Emax$ using the attractive Hubbard model. The density of states is given in units of $1/t$ with $t$ the hopping energy. On panel \textbf{c}, we show the mean density of states of the numerical model (black line) along with the normalized variance $\sigma^2$ (red). We plot analytical prediction from Eq. \eqref{LDOS_corr_main_text} for $g\in\left\{10,11,12\right\}$. We stress that no free parameter has been optimized. On panel \textbf{d}, we plot the LDOS distribution at fixed energy along with the corresponding gaussian (dotted line) and log-normal (solid line) distributions (\textit{ie:} of same variance). Distribution functions are given in Appendix \ref{Sec:AppC}. On panel \textbf{e}, we show the RMS deviation to respectively gaussian (dark grey) and log normal (light grey) models. Panels \textbf{f,g} are analogous to \textbf{d,e} for the experimental dI/dV distribution.}
\label{Figure5}
\end{figure*}

Here, we study the distribution function of both our experimental $dI/dV$ maps and the ones obtained from the attractive Hubbard model. Starting with the numerical results,  we plot on Figure \ref{Figure5}\pf{\textbf{d}}, the distribution of the LDOS distribution at fixed energy along with the corresponding (i.e. of same variance) Gaussian and log-normal laws (see Appendix \ref{Sec:AppC}). A clear log-normal behavior is identified on all the energy range between $\Emax$ and  $3\Emax$ as confirmed on panel \ref{Figure5}\pf{\textbf{e}} which compares the root-mean-squared (RMS) deviation of the distribution to either the Gaussian or the log-normal model. Turning to the experimental data, we show on Figure \ref{Figure5}\pf{\textbf{f}}, the $dI/dV$ distribution with Gaussian (dotted line) and log-normal laws (solid line) of same variance at voltage corresponding to $\Emax$ and $1.5 \Emax$. Like before, we plot on panel \ref{Figure5}\pf{\textbf{h}} the RMS deviation to the Gaussian or log-normal model. The log-normal model being more accurate is a hint of the multifractal nature of the LDOS fluctuations close to the superconducting coherence peak.

\section*{Conclusion}
As a model system for two-dimensional weakly disordered superconductor where multifractal superconductivity is being actively pursued, we prepared a single layer of lead on Si(111) where electrons are anti-localized in a controlled crystalline disorder and become superconducting below 1.8 K. Using scanning tunneling spectroscopy, we report the measurement of tunneling conductance fluctuations with exquisite spatial and spectral resolutions on scales exceeding the superconducting coherence length. To our knowledge, this study is the first to link experimentally spatial LDOS fluctuations with weak \CORR{anti-localization} physics. To support our analysis, we used two theoretical approaches, a semi-analytical and a numerical one. Our numerical approach consisted in an attractive Hubbard model with a disorder level tuned to match the experiment and probed \Mathieu{both gap-width and LDOS} fluctuations close to the superconducting coherence peak. Our experimental, semi-analytical and numerical results are shown to be quantitatively consistent with the mesoscopic fluctuations physics in the weakly \CORR{anti-}localized regime usually probed through transport measurements. The LDOS fluctuation's amplitude depends on two local parameters which can be probed and quantitatively extracted in this way: the metal's conductance and an effective electronic dephasing rate. 


\Igor{
\begin{acknowledgements}
This work was supported by the French national research fund managed by the ANR through the project RODESIS having the contract number ANR-16-CE30-0011-01. The work of I.S.B. was funded in part by the Russian Ministry of Science and Higher Educations, the Basic Research Program of HSE, and by the Russian Foundation for Basic Research, grant No. 20-52-12013.
\end{acknowledgements}
}

\appendix

\section{Mean density of state\label{Sec:AppA}}

\begin{figure*} 
\centering
\includegraphics[width=18cm]{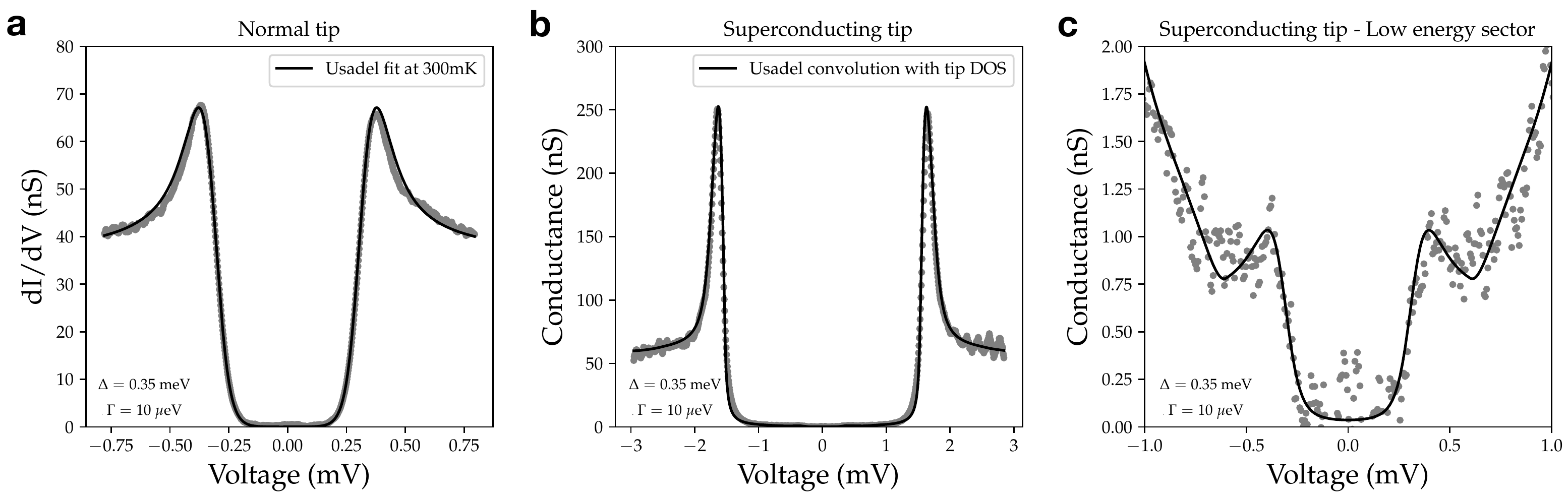} 
\caption{\textbf{Fit of the mean conductance} On \textbf{a}, the mean differential conductance measured with a platinum tip is shown as a function of the bias voltage along with the best Usadel fit. On panels \textbf{b}, we show the mean dI/dV curve of the large scale spectroscopic map measured with a bulk lead tip while the solid line is a convolution of the Usadel density of states for the sample with the tip density of states at 300 mK. \Mathieu{On panel \textbf{c}, we show a zoom in the low energy sector for the superconducting tip allowing a very precise extraction of $\Delta_{SIC}$ and $\Gamma_{SIC}$ (\textit{cf:} excellent agreement of the Usadel model with the experiment).}}
\label{figure:Fit}
\end{figure*}

We use the Usadel model \cite{usadel_generalized_1970,gueron_quasiparticles_nodate} to describe the diffusive SIC phase as the mean free path is much smaller than the superconducting coherence length in this sample: $\ell {\sim} 1$ nm $\ll \xi {\sim} 50$ nm. In more details, we use the spectral angle $\theta_E$ parametrisation, with $\theta_E$ a solution of an homogenous Usadel equation with a depairing term~$\Gamma$:

\begin{gather}
iE \sin(\theta_E)+\Delta \cos(\theta_E) - \Gamma \sin(\theta_E)\cos(\theta_E)=0
\label{appendixA_usadel} 
\end{gather}
The solution of this equation $\theta_E$ yields $X_E = 
\cos(\theta_E)$ ($=\frac{E}{\sqrt{E^2-\Delta^2}}$ at $\Gamma=0$) from which the density of states $\re (X_E)$ is obtained.

Using Eq. \eqref{appendixA_tunneling_current} and Eq. \eqref{appendixA_usadel}, we convolve the density of states of the tip with the one of the sample (along with the Fermi distribution at 300 mK) in order to reproduce the mean differential conductance.
\begin{gather}
I \propto \int \rho_{\rm tip}(E-eV) \rho_{\rm SIC}(E,\bm{r})(f(E)-f(E-eV)) dE
\label{appendixA_tunneling_current}
\end{gather}
where $\rho_{\rm tip}$ and $\rho_{\rm SIC}$ are the respective energy-dependent DOS of the tip and the SIC phase. As detailed on Figure~\ref{figure:Fit}, we find that the lead tip is very well described by a Usadel superconductor ($\DeltaTip = 1.345$ meV, $\GammaTip = 20 \mu$eV). The SIC phase is found to be very well described by $\DeltaSIC=0.35$ meV and $\GammaSIC = 10 \mu$eV in excellent agreement with additional measurements using a normal tip (Figure \ref{Figure1}\pf{\textbf{b}} and \ref{figure:Fit}\pf{\textbf{a}}) and with earlier works~\cite{brun_remarkable_2014,cherkez_proximity_2014}.




\begin{widetext}

\section{Differential conductance variance computation\label{Sec:AppB}}

We now attempt to rationalise the tunneling conductance spatial fluctuations. Considering a simplified expression for the tunneling current, we write it as a zero temperature convolution of tip and sample density of states Eq.\eqref{appendixA_tunneling_current}. Experimentally, the tip's height above the sample and thus the transmission's coefficient $t$ is not constant but rather controlled by fixing the high voltage current $I(V_\Lambda=$ 3 mV) to 20 pA. Trying to estimate density of state leads us \CORR{to compute}: 
\begin{equation}
\eta(V,\bm{r})= \frac{\overline{I(V_\Lambda)}}{I(V_\Lambda,\bm{r})} \frac{\partial I}{\partial V}(V,\bm{r})
\label{eq:eta}
\end{equation}

\CORR{At $T=0$, we write for the tunneling current $I$ and the tunneling conductance $dI/dV$:
\begin{gather}
I =  |t|^2 \int\limits_{0}^{V} dE \rho_{\rm tip}(E-eV) \rho(E,\bm{r}) \quad \textrm{ and } \quad \frac{\partial I}{\partial V} = 
|t|^2 \rho_{\rm tip}(0) \rho(V,\bm{r})-|t|^2 \int\limits_0^{V} dE  \rho^\prime_{\rm tip}(E-V) \rho(E,\bm{r})
\label{appendixB_tunneling_current}
\end{gather}

where $\rho^\prime_{\rm tip}(E) \equiv d\rho_{\rm tip}(E)/dE$. It is convenient to introduce the following notations
\begin{gather}
\bar{j}(V) = \int\limits_0^{V} dE \rho_{\rm tip}(E-V)\overline{\rho}(E) \quad \textrm{ and }\quad \bar{j}^\prime(V) =\rho_{\rm tip}(0)\bar\rho(V)-\int\limits_0^{V} dE \rho^\prime_{\rm tip}(E-V) \bar \rho(E) .
\end{gather}
Assuming that fluctuations $\delta \rho(E,\bm{r}) = \rho(E,\bm{r}) - \bar{\rho}(E)$ near the average DOS $\bar{\rho}(E)$ are weak, we find:
\begin{gather}
\frac{\delta \eta(V,\bm{r})}{\overline{I(V_\Lambda)}}  = \frac{\rho_{\rm tip}(0) \delta \rho(V,\bm{r})}{\bar{j}(V_\Lambda)} - \int\limits_0^{\infty} dE  \, \delta \rho(E,\bm{r}) 
\frac{\rho_{\rm tip}^\prime(E-V) \bar{j}(V_\Lambda)\Theta(V-E) + \rho_{\rm tip}(E-V_\Lambda) \bar{j}^\prime(V)\Theta(V_\Lambda-E)}{[\bar{j}(V_\Lambda)]^2} .
\label{eq:7}
\end{gather}
Here $\Theta(x)=1$, for $x>0$ and zero otherwise.} Hence we obtain
\begin{align}
\frac{\langle \delta \eta(V_1,\bm{r}_1)\delta \eta(V_2,\bm{r}_2)\rangle}{[\overline{I(V_\Lambda)}]^2} & =
\frac{\rho_{\rm tip}^2(0)}{\bar{j}(V_\Lambda)\bar{j}(V_\Lambda)}\langle \delta \rho(V_1,\bm{r}_1)\delta \rho(V_2,\bm{r}_2)\rangle
\notag \\
- & \frac{\rho_{\rm tip}(0)}{\bar{j}(V_\Lambda)}  \int\limits_0^{\infty} dE  \, \langle \delta \rho(V_1,\bm{r}_1) \delta \rho(E,\bm{r}_2) \rangle \frac{\rho_{\rm tip}^\prime(E-V_2) \bar{j}(V_\Lambda)\Theta(V_2-E) + \rho_{\rm tip}(E-V_\Lambda) \bar{j}^\prime(V_2)\Theta(V_\Lambda-E)}{[\bar{j}(V_\Lambda)]^2}\notag \\
- & \frac{\rho_{\rm tip}(0)}{\bar{j}(V_\Lambda)}  \int\limits_0^{\infty} dE  \, \langle \delta \rho(V_2,\bm{r}_2) \delta \rho(E,\bm{r}_1) \rangle \frac{\rho_{\rm tip}^\prime(E-V_1) \bar{j}(V_\Lambda)\Theta(V_1-E) + \rho_{\rm tip}(E-V_\Lambda) \bar{j}^\prime(V_1)\Theta(V_\Lambda-E)}{[\bar{j}(V_\Lambda)]^2}
\notag \\
+ & \int\limits_0^{\infty} dE_1 \int\limits_0^{\infty} 
dE_2\, \langle \delta \rho(E_1,\bm{r}_1)
\delta \rho(E_2,\bm{r}_2)\rangle 
\frac{\rho_{\rm tip}^\prime(E_1{-}V_1) \bar{j}(V_\Lambda)\Theta(V_1{-}E_1) {+} \rho_{\rm tip}(E_1{-}V_\Lambda) \bar{j}^\prime(V_1)\Theta(V_\Lambda{-}E_1)}{[\bar{j}(V_\Lambda)]^2}
\notag \\
& \times
\frac{\rho_{\rm tip}^\prime(E_2{-}V_2) \bar{j}(V_\Lambda)\Theta(V_2{-}E_2) + \rho_{\rm tip}(E_2{-}V_\Lambda) \bar{j}^\prime(V_2)\Theta(V_\Lambda{-}E_2)}{[\bar{j}(V_\Lambda)]^2} .
\label{appendixB_dos}
\end{align}

\section{Iso-energy LDOS distribution\label{Sec:AppC}}
We compare the iso-energy dI/dV distributions to either Gaussian or log-normal models of variance $\sigma$. The distribution of the logarithm of normalized density of states $x=\textrm{log}(\rho(E,r)/\langle\rho(E,r)\rangle_r)$ writes for a Gaussian distribution: 
\begin{equation}
   f_\textrm{log-n}(x) = \frac{1}{\sqrt{2\pi \sigma^2}} \textrm{exp}\Big[x-\frac{(e^x-1)^2}{2\sigma^2}\Big]
\end{equation}

For a log normal distribution which is expected by the \Igor{\sout{sigma-model}theoretical} analysis and recovered in the numerical work, the log of the normalized LDOS $x$ is distributed following:
\begin{equation}
     f_\textrm{n}(x) = \frac{1}{\sqrt{2\pi \sigma^2}}\textrm{exp}\Big[\frac{-(x+\sigma^2/2)^2}{2\sigma^2}\Big]
\end{equation}

\section{Gap width fluctuations \label{Sec:AppD}}

Let us assume that LDOS $\rho(E)$ at a given realization of disorder potential has the single maximum (for positive energies) at $E_{\rm max}$ as a function of energy. Provided deviation $\delta E_{\rm max}=E_{\rm max}-\overline{E}_{\rm max}$ from the energy $\overline{E}_{\rm max}$ of the maximum in the average LDOS $\overline{\rho}(E)$, we can write 
\begin{gather}
0=\rho^\prime(E_{\rm max}) \simeq   \bar{\rho}^\prime({E}_{\rm max})  
+ \delta{\rho}^\prime({E}_{\rm max}) \simeq \bar{\rho}^\prime(\bar{E}_{\rm max})+
\bar{\rho}^{\prime\prime}(\bar{E}_{\rm max}) \delta E_{\rm max}
+\delta{\rho}^\prime({E}_{\rm max})
\simeq\bar{\rho}^{\prime\prime}(\bar{E}_{\rm max}) \delta E_{\rm max}
+\delta{\rho}^\prime({E}_{\rm max}),\notag \\
\quad \Longrightarrow \quad 
\overline{(\delta E_{\rm max})^2} \simeq \frac{\overline{(\delta{\rho}^\prime(\bar{E}_{\rm max}}))^2}{(\bar{\rho}^{\prime\prime}(\bar{E}_{\rm max}))^2}
 .
\end{gather}
Here the variance of the energy derivative of the LDOS can we read from Eq. 
\eqref{LDOS_corr_main_text}

The quantity $X_E$ for the Usadel equation \eqref{appendixA_usadel}, near $E_{\rm max}$ can be approximated as (assuming $\Gamma\ll\Delta$) \cite{abrikosov1960contribution,fominov2016subgap}

\begin{equation}
X_E = \left (\frac{\Delta}{\Gamma}\right)^{1/3} 
f\left (\frac{E-E_g}{\Gamma^{2/3}\Delta^{1/3}}\right ) ,
\end{equation}
where the spectral gap $E_g=\Delta(1-3 (\Gamma/\Delta)^{2/3}/2)$. Then we find the following estimates
\begin{gather}
\overline{(\delta{\rho}^\prime(\bar{E}_{\rm max}}))^2
\simeq \frac{c_1}{4\pi g} \frac{\Delta^2}{\Gamma^2}, \qquad 
\bar{\rho}^{\prime\prime}(\bar{E}_{\rm max}) \simeq c_2 \left (\frac{\Delta}{\Gamma}\right )^{5/3}, \qquad \overline{(\delta E_{\rm max})^2} \simeq \frac{c}{g} \left (\frac{\Gamma}{\Delta}\right )^{4/3},
\end{gather}
where $c_1\approx 0.23$, $c_2\approx-0.24$, and $c=c_1/(4\pi c_2^2)\approx 0.32$.

\referee{\section{Symmetry with respect to the Fermi energy \label{Sec:AppE}}
Our analytical calculations predict the same normalized variance in the negative and positive energy range. It is important for the consistency of our analysis to check this symmetry in the experiments. On Figure \ref{figure:negative_bias}, we show that not only the visual aspect of LDOS map at the coherence peak but also the energy-dependent normalized LDOS standard deviation is perfectly symmetric with respect to the Fermi level.

\begin{figure*} 
\centering
\includegraphics[width=18cm]{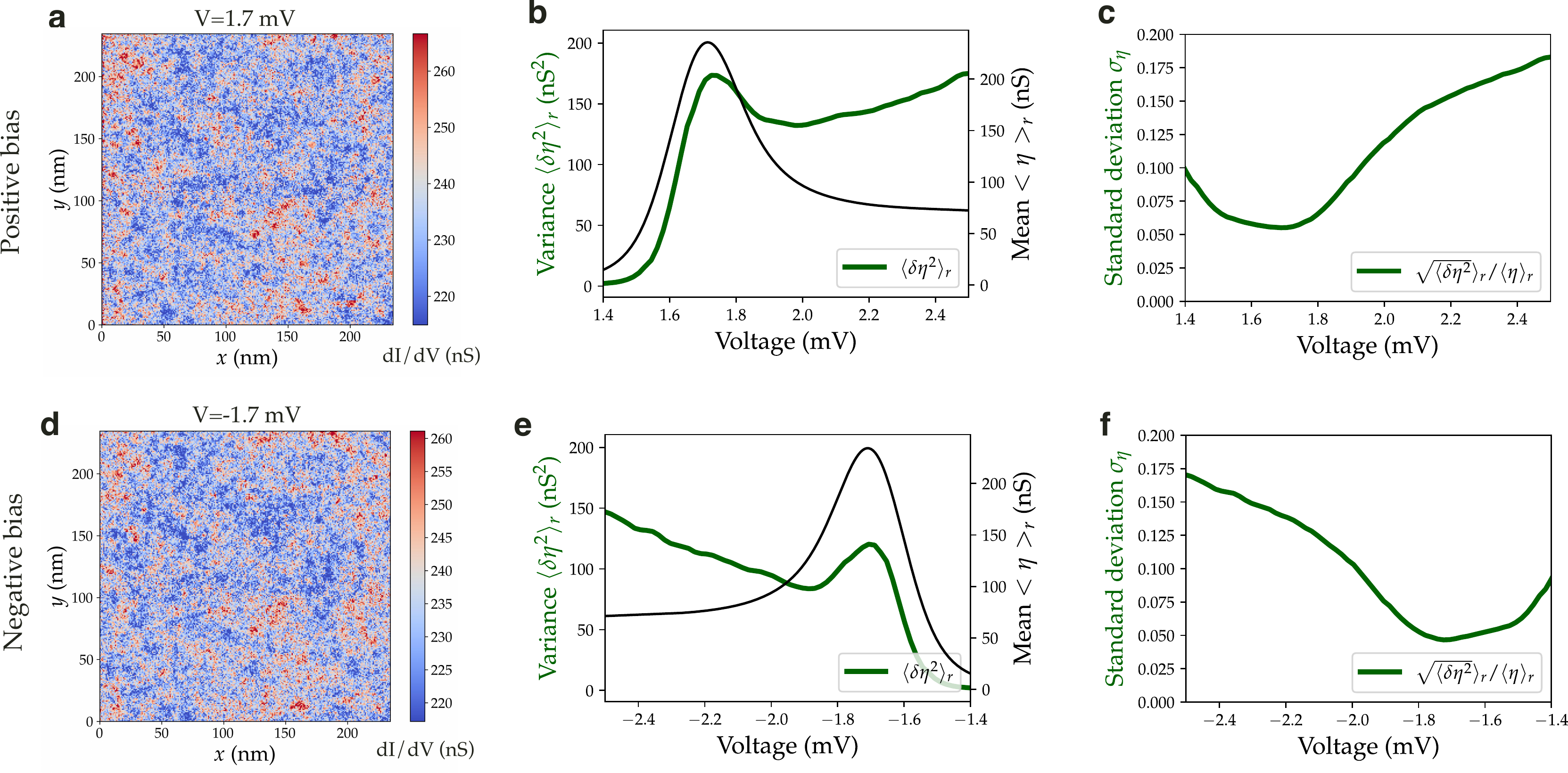} 
\caption{\referee{\textbf{Symmetry with respect to Fermi energy} For positive bias voltage, we show the LDOS map at $\Emax=1.7$ mV (panel \textbf{a}), the variance and mean value of tunneling conductance (panel \textbf{b}) and the normalized standard deviation $\sigma_\eta$ (panel \textbf{c}). On panels \textbf{d-f}, we show the corresponding data for negative bias voltage.}}
\label{figure:negative_bias}
\end{figure*}}




\end{widetext}

\bibliography{disorder}
\end{document}